# Complementarity and uncertainty in a two-way interferometer

Jie-Hui Huang,[1] Shi-Yao Zhu,[1,2]
[1]Department of Physics, Hong Kong Baptist University, Hong Kong, China
[2]Key Laboratory of Quantum Information,
University of Science and Technology of China, Hefei, 230026, China

## Abstract

In terms of operator, the two complementary quantities, the *predictability* and *visibility*, are reinvestigated in a two-way interferometer. One Hermitian operator and one non-Hermitian operator (composed of two Hermitian operators) are introduced for the *predictability* and *visibility*, respectively. The *predictability* and *visibility* can not be measured exactly simultaneously, due to the non-commutation between the two operators. The sum of the variances of the *predictability* and *visibility* (the total variance), is used to measure the uncertainty, which is linked to the complementarity relation through the equation, $(\Delta P)^2 + (\Delta V_f)^2 + P^2 + V^2 = 2$. This new description for the *predictability* and *visibility* connects the complementarity and the uncertainty relations, although neither of them can be derived directly from the other.

## I. INTRODUCTION

The uncertainty principle and complementarity are at the conceptual heart of the quantum theory. The complementarity emphasizes equally real but mutually exclusive properties, such as the wave-particle duality. For two complementary properties, the observation of either one will preclude the observation of the other. The uncertainty principle concerns the fluctuations in the quantum mechanics and predicts the limit of precision for measuring two quantities simultaneously. Since the uncertainty principle and complementarity are both associated with quantum measurements, it is natural to ask: what is the relationship between the two principles? Scully and co-workers studied this issue by proposing a quantum eraser scheme in an atom interferometer [1-6], where the loss and restoration of interference pattern can be explained through correlations between detector and atomic motion, rather than random momentum kicks. Since the standard position-momentum uncertainty relation plays no role in describing the wave-particle duality in this proposed atom interferometer and later experimental realizations [7-12], it is concluded that "the principle of complementarity is much deeper than the uncertainty relation" [13]. However, Storey *et al.* argued that the repeated emission and reabsorption of microwave photons by the atom actually indicates the momentum kicks in the above quantum eraser scheme. Therefore, they believed that "the principle of complementarity is a consequence of the Heisenberg uncertainty relation" [14-16].

Another way to investigate the relationship between the two principles is to quantify the complementary properties appropriately, and compare the results with the uncertainties in the same system. In 1979, the "wavelike" and "particlelike" properties of light in Young's double-slit experiment was first quantified, by Wootters and Zurek, as the sharpness of the interference pattern and the amount of information on the photons' trajectories [17], respectively. In a similar manner, Glauber *et al.* achieved a complementarity relation between the "particle knowledge" *P* and the "wave knowledge" *W* in a two-way interferometer [18-20], which can be regarded as the quantification of wave-particle duality in this particular system.

Although the derivation of complementarity relation usually does not make use of uncertainty relation in any form, some connections between them can be established [21-29]. For example, Björk *et al.* presented an explicit expression to link the *predictability P*, *visibility V* and a "normalized uncertainty product" of two hermitian operators $\hat{A}$ and $\hat{B}$. Here the first hermitian operator $\hat{A}$ can be arbitrarily chosen, and the second one, $\hat{B}$, is complementary to $\hat{A}$, with its eigenstates being equally weighted superpositions of the eigenstates of $\hat{A}$ [21]. In Refs.[22, 23], the *predictability P* and *visibility V* are simply represented by two Hermitian operators $\sigma_z$ and $\sigma_x$. The Heisenberg-type uncertainty of these two Hermitian operators (product of the variances of $\sigma_z$ and $\sigma_x$) directly lead to a "complementarity relation". However, this "complementarity relation" depends on three operators, not only $\sigma_z$ and $\sigma_x$, but also $\sigma_y$. In fact, the measurement of $\sigma_x$ does not provide accurate prediction for the fringe *visibility* in a general case, which means the single operator $\sigma_x$ is not enough to represent the *visibility V* in a two-way interferometer.

In this paper we introduce one Hermitian operator and one non-Hermitian operator, composed of two Hermitian operators, to represent the *predictability* and *visibility* in a two-way interferometer, respectively. The variance of a non-Hermitian operator is evaluated through the error propagation method [30]. Besides the standard Heisenberg-type uncertainty (product of two variances), the total variance (sum of two variances) is also used to describe the uncertainty in the interferometer. By describing the *predictability* and *visibility* in this way, a direct connection between the duality relation and the uncertainty relation can be established.

## II. COMPLEMENTARITY RELATIONS AND FLUCTUATION RELATIONS IN A TWO-WAY INTERFEROMETER

In a two-way interferometer illustrated in Fig. 1, the initial state of a photon, which is emitted from a light source (LS) and enters this interferometer from left to right, can be generally described by

$$\rho^{(i)} = \frac{1 + s^{(i)} \cdot \sigma}{2} = \frac{1}{2}(1 + s_x^{(i)}\sigma_x + s_y^{(i)}\sigma_y + s_z^{(i)}\sigma_z), \tag{1}$$

where $\sigma = (\sigma_x, \sigma_y, \sigma_z)$ are Pauli's spin operators and $s^{(i)} = tr\{\sigma\rho^{(i)}\}$ is the initial Bloch vector with $(s_x^{(i)})^2 + (s_y^{(i)})^2 + (s_z^{(i)})^2 \leq 1$ (the upper bound corresponds to pure states). The phase shifter (PS) introduces a phase difference $\phi$ between the "+" and "–" paths through the operation $S_1 = \exp(i\frac{\phi}{2}\sigma_z)$. The beam merger (BM), which is characterized by the operator $S_2 = \exp(i\frac{\pi}{4}\sigma_y)$ [31-33], combines the light beams from the two paths. The phase shifter (S1) and the beam merger (S2) turn the initial state (1) to the final state

$$\rho^{(f)} = S_2^+ S_1^+ \rho^{(i)} S_1 S_2 = \frac{1 + s^{(f)} \cdot \sigma}{2}, \tag{2a}$$

with

$$s^{(f)} = (s_z^{(i)}, s_x^{(i)}\sin\phi + s_y^{(i)}\cos\phi, s_y^{(i)}\sin\phi - s_x^{(i)}\cos\phi). \tag{2b}$$

The difference of the probabilities for taking "+" and "−" ways give the photon's which-way information, i.e. *predictability*,

$$P = \left|tr\{\sigma_z \rho^{(i)}\}\right| = \left|<\sigma_z>\right| = \left|s_z^{(i)}\right|. \tag{3}$$

The fringe *visibility* of the interference pattern, which can be observed by recording the detection probability with varying phase, is determined by the maximum and minimum values of the probability distribution, read as

$$V = \frac{p_{max} - p_{min}}{p_{max} + p_{min}} \tag{4}$$

Due to the one-to-one correspondence between the probability $p$ of detecting a photon in the "+" way and the expectation value of the operator $\sigma_z$, $p = \frac{1}{2}(<\sigma_z> + 1)$, the *visibility* (4) can be rewritten as

$$V = \frac{<\sigma_z>'_{max} - <\sigma_z>'_{min}}{2 + <\sigma_z>'_{max} + <\sigma_z>'_{min}} = \frac{(s_z^{(f)})_{max} - (s_z^{(f)})_{min}}{2 + (s_z^{(f)})_{max} + (s_z^{(f)})_{min}}, \tag{5}$$

where the prime indicates that the averaging is operated on the final photonic state $\rho^{(f)}$. Since the maximum and minimum values of $s_z^{(f)}$ in the Bloch vector $s^{(f)}$ satisfy,

$$(s_z^{(f)})_{max} = -(s_z^{(f)})_{min} = \sqrt{(s_x^{(i)})^2 + (s_y^{(i)})^2}. \tag{6}$$

the fringe *visibility* (4) is finally expressed as a function of the x- and y-components of the initial Bloch vector $s^{(i)}$,

$$V = \sqrt{(s_x^{(i)})^2 + (s_y^{(i)})^2}. \tag{7}$$

Noting that $s_z^{(f)}$ and $s_z^{(f)}$ correspond to the real and imaginary parts of the off-diagonal element of the 2×2 density matrix (1), respectively. The above formula is equivalent to the result, $V = 2|\rho_{21}|$, in Refs. [34, 35]. The combination of Eqs.(3) and (7) leads to the complementarity relation [19, 20]:

$$P^2 + V^2 = (s_x^{(i)})^2 + (s_y^{(i)})^2 + (s_z^{(i)})^2 \leq 1, \tag{8}$$

where the equal sign holds for all the pure states.

Based on the results (3) and (7), now we introduce two operators $\hat{P}$ and $\hat{V}$, defined by

$$\hat{P} = \sigma_z \tag{9a}$$

to represent the *predictability* of the which-way information, and

$$\hat{V} = \sigma_x + i\sigma_y, \tag{9b}$$

a non-Hermitian operator (composed of two Hermitian operators), to represent the fringe *visibility*. Just as indicated in the definition (4), the usual way to calculate the fringe *visibility* is to find the maximum and minimum detection probabilities, which means the probability should be measured with respect to every phase shift in the range $[0, 2\pi]$ in principle. The operator (9b) implies that the *visibility* can also be measured on the initial state and only two measurements (on $\sigma_x$ and $\sigma_y$) are required in experiments.

As well known, it is impossible to assign sharp values to the *predictability* and the fringe *visibility* simultaneously [23], that is to say, if the *predictability* and the fringe *visibility* are represented by two operators, they should not commute with each other. This condition is satisfied by the two operators in (9), because $[\hat{P}, \hat{V}] = 2\hat{V}$ (more explicitly, $[\hat{P}, \sigma_x] = 2i\sigma_y$ and $[\hat{P}, \sigma_y] = -2i\sigma_x$). On the other hand, the non-commutation between the two operators in (9) is in accord with the mutually exclusive property of the "wavelike" and "particlelike" behavior. Since the complementarity relation and the uncertainty relation of the *predictability* and fringe *visibility* both are related to the non-commutation between the two operators, it is an evidence that the complementarity and the uncertainty principle are not completely independent.

Now let us consider the uncertainty in the interferometer. According to the definitions (9), the *visibility* is determined by the measurements on $\sigma_x$ and $\sigma_y$, while the *predictability* is determined by $\sigma_z$. Based on the definition $(\Delta \hat{A})^2 = <\hat{A}^2> - (<\hat{A}>)^2$, the variances of the three operators $\sigma_x$, $\sigma_y$ and $\sigma_z$ are easy to obtain,

$$(\Delta \sigma_x)^2 = 1 - (<\sigma_x>)^2 = 1 - (s_x^{(i)})^2 \tag{10a}$$

$$(\Delta \sigma_y)^2 = 1 - (<\sigma_y>)^2 = 1 - (s_y^{(i)})^2 \tag{10b}$$

$$(\Delta \sigma_z)^2 = 1 - (<\sigma_z>)^2 = 1 - (s_z^{(i)})^2 \tag{10c}$$

Noting that two different states with opposite expectation values have the same variance, i.e., $(\Delta \sigma_u)^2 = (\Delta \sigma_u)'^2$ if $<\sigma_u> = -<\sigma_u'>$ $(u = x, y, z)$, we can get the variance for the *predictability*,

$$(\Delta P)^2 = (\Delta \sigma_z)^2 = 1 - (<\sigma_z>)^2 = 1 - P^2 = 1 - (s_z^{(i)})^2. \tag{11}$$

The two results (3) and (11) imply that high *predictability* always corresponds to small variance.

The situation is a little more complicated when concerning the fringe *visibility*, which is represented by a non-Hermitian operator (9b). Because $\hat{V}$ is a non-hermitian operator, its variance can not be directly evaluated in the above way. Since the *visibility* V is determined by the measurements of $\sigma_x$ and $\sigma_y$ through the relation $V = |<\hat{V}>| = \sqrt{(<\sigma_x>)^2 + (<\sigma_y>)^2}$, its fluctuation depends on the variance and and co-variance of $\sigma_x$ and $\sigma_y$. By regarding $\sigma_x$ and $\sigma_y$ as two variables and using the error propagation method [30], we can calculate the variance of V through

$$(\Delta V_f)^2 = (\frac{\partial V}{\partial \sigma_x})^2 (\Delta \sigma_x)^2 + (\frac{\partial V}{\partial \sigma_y})^2 (\Delta \sigma_y)^2 + 2 \frac{\partial V}{\partial \sigma_x} \frac{\partial V}{\partial \sigma_y} (\Delta \sigma_x \sigma_y)$$

$$= (\frac{<\sigma_x>}{V})^2 (\Delta \sigma_x)^2 + (\frac{<\sigma_y>}{V})^2 (\Delta \sigma_y)^2 + 2 \frac{<\sigma_x>}{V} \frac{<\sigma_y>}{V} (\Delta \sigma_x \sigma_y) \tag{12}$$

$$= 1 - V^2.$$

Here $\Delta \sigma_x \sigma_y$ represents the co-variance between the two operators $\sigma_x$ and $\sigma_y$, which is [36]

$$\Delta\sigma_x\sigma_y = \frac{1}{2}<\sigma_x\sigma_y + \sigma_y\sigma_x> - <\sigma_x><\sigma_y> = -s_x^{(i)}s_y^{(i)}. \qquad (13)$$

With the two variances (11) and (12), we now have two basic ways to describe the uncertainty of this system. The first one is to calculate the product of the two variances, i.e,

$$(\Delta P)^2(\Delta V_f)^2 = (1-P^2)(1-V^2) \in [0,1]. \qquad (14a)$$

This uncertainty equation does not show the direct connection with the complementarity relation (8), as depends on $P^2 + V^2$ and also $P^2V^2$. Therefore, the uncertainty equation is not a logical consequence of the complementarity, and *vise versa*.

Alternatively, we choose the sum of the two variances to describe the uncertainty, which leads to,

$$(\Delta P)^2 + (\Delta V_f)^2 = (1-P^2) + (1-V^2). \qquad (14b)$$

We can rewrite Eq.(14b) as

$$(\Delta P)^2 + (\Delta V_f)^2 + P^2 + V^2 = 2. \qquad (15)$$

This equation is our main result, where the uncertainty (total variance) is directly connected to the complementary relation. This is the link between the complementarity and the uncertainty principle, although neither of them can be derived directly from the other. Since the sum of the total variance and the total information of the *predictability* and the fringe *visibility* is a constant, the increasing (decreasing) of the total information $P^2 + V^2$ always results in the linear decreasing (increasing) of the uncertainty $(\Delta P)^2 + (\Delta V_f)^2$. In other words, the more useful information, the smaller fluctuation. The uncertainty has the minimum of 1 with the highest total information of the *predictability* and the fringe *visibility* $P^2 + V^2 = 1$. The upper bound of the total variance (maximum uncertainty) is reached when $s_x^{(i)} + s_y^{(i)} + s_z^{(i)} = 0$ and no useful which-way information and fringe *visibility* information is available in this special case.

Recently, Busch and Shilladay investigate the duality and uncertainty relation in the same interferometer by measuring the "path contrast" (*predictability*) and the "interference contrast" (*visibility*) through two hermitian operators $\sigma_z$ and $\sigma_x$ [23]. From the variances of the two operators, $(\Delta\sigma_z)$ and $(\Delta\sigma_x)$, and the normally defined uncertainty $(\Delta\sigma_z)(\Delta\sigma_x)$, they find the complementarity relation, $P^2 + V_x^2 + V_y^2 (= P^2 + V^2) \leq 1$. However, this complementarity relation depends on three operators $\sigma_x$, $\sigma_y$ and $\sigma_z$ (or as emphasized one path and two complementary interference observables), instead of the two operators for the uncertainty. Therefore, this complementarity equation is not directly connected with the uncertainty equations.

Both the complementarity and the uncertainty are derived from the two non-commuting operators ($\hat{P}$ and $\hat{V}$). The complementarity (the total available information of the which-way *predictability* and the fringe *visibility* in the two-way interferometer) is determined by Eq.(8), while the uncertainty (defined by the total variance) is determined by Eq.(15). The sum of the complementarity (the total information of the *predictability* and the fringe *visibility*) and the uncertainty (the total variance) is a constant, which gives us the link between the complementarity and the uncertainty, although one can not be derived directly from the other.

It is interesting to ask what happens if the photon discussed above is entangled with another one. For example, if the two-way interferometer in Fig.1 is a part (left-side or right-side) of a typical two-particle four-beam interferometer [37, 38] shown in Fig.2, then the two subsystems, i.e. the single particle 1 (entering the two left beams) and 2 (entering the two right beams), are described by two reduced density matrices,

$$\rho_1^{(i)} = tr_2\{\rho_{1,2}^{(i)}\} = \begin{pmatrix} |a|^2 + |b|^2 & ac^* + bd^* \\ a^*c + b^*d & |c|^2 + |d|^2 \end{pmatrix}, \tag{16a}$$

and

$$\rho_2^{(i)} = tr_1\{\rho_{1,2}^{(i)}\} = \begin{pmatrix} |a|^2 + |c|^2 & ab^* + cd^* \\ a^*b + c^*d & |b|^2 + |d|^2 \end{pmatrix}, \tag{16b}$$

where an initial pure state

$$\rho_{1,2}^{(i)} = \begin{pmatrix} a \\ b \\ c \\ d \end{pmatrix}(a^* \quad b^* \quad c^* \quad d^*), \quad \text{with} \quad |a|^2 + |b|^2 + |c|^2 + |d|^2 = 1 \tag{17}$$

is assumed for the two-particle composite system. By applying the two operators (9a) and (9b) we introduced for $\hat{P}$ and $\hat{V}$, it is easy to get the *predictability* and *visibility* of the two single-particle subsystems,

$$\begin{aligned} P_1 &= \left||a|^2 + |b|^2 - |c|^2 - |d|^2\right|; \\ P_2 &= \left||a|^2 + |c|^2 - |b|^2 - |d|^2\right|, \end{aligned} \tag{18a}$$

and

$$\begin{aligned} V_1 &= 2|a^*c + b^*d|; \\ V_2 &= 2|a^*b + c^*d|, \end{aligned} \tag{18b}$$

The two correlated particles in the pure state (17) have the same total information of the "wavelike" and "particlelike" behavior, that is

$$|P_1|^2 + |V_1|^2 = |P_2|^2 + |V_2|^2 = 1 - 4|ad - bc| = 1 - C^2. \tag{19}$$

The total variance of *predictability* and *visibility* (15) of particle $k$ ($k = 1; 2$) in this special system can be calculated as,

$$(\Delta P)_k^2 + (\Delta V_f)_k^2 = 2 - (s_x^{(i)})_k^2 - (s_y^{(i)})_k^2 - (s_z^{(i)})_k^2 = 1 + C^2. \tag{20}$$

Here $C = 2|ad - bc|$ is the entanglement (concurrence) [39] of the two correlated particles in the pure state (17). As shown in Eqs.(19) and (20), the total available single-particle information and the °uctuation (total variance) in this two-particle composite system both are determined by the entanglement (concurrence) $C$. An entanglement between the two particles always results in decreasing of the single-particle information and causes uncertainty increasing. In the extreme case of $C = 1$, which means the two particles are maximally entangled to each other, there is no single-particle's "wavelike" and "particlelike" information at all, because $|P_k|^2 + |V_k|^2 = 0 (k = 1, 2)$ in this case according to Eq. (19). At the same time, $C = 1$ also means the maximum uncertainty in this composite system, $(\Delta P)_k^2 + (\Delta V_f)_k^2 = 2$. The other extreme case, $C = 0$, actually means that the two particles are separable to each other, thus we can investigate the two subsystems separately and the main results are presented in Eqs.(8), (14a) and (15). The relation (19) is called *wave-particle-entanglement* "triality" relation [40] by regarding the concurrence as the third equally real but mutually exclusive quantity between two particles, besides the "wavelike" and "particlelike" properties of single particle. However, the entanglement is a kind of uncertainty [23] for the single-particle system (an extra variance compared with a pure single-particle system) as shown by Eq.(20). In fact, Eqs. (19) and (20) can be considered to be included in the complementarity and uncertainty equation (15) based on the two operators $\hat{P}$ and $\hat{V}$ for the single particle system.

## III. CONCLUSIONS

In this paper we defined a Hermitian operator (9a) for the *predictability* and a non-Hermitian operator (composed of two Hermitian operators) (9b) for the fringe *visibility* in a two-way interferometer as the two operators of the complementarity. The complementarity relation is derived, and the variances of the two quantities are evaluated. Using the total variance (sum of the two variances instead of the variance product) of the two operators, the direct connection between the complementarity relation and the uncertainty relation (Eq.15) is established. The complementary relation (8) limits the total available information of the *predictability* and *visibility*, while the uncertainty relations (14a) and (14b) predict the range of the precision of the *predictability* and *visibility*. Equation (15), $(\Delta P)^2 + (\Delta V_f)^2 + P^2 + V^2 = 2$, is the direct link between the complementarity and the uncertainty principle, though neither of them can be derived directly from the other.

**Acknowledgments**

This research was supported by RGC Grant No. (NSFC05-06.01) of HK Government and FRG of HK Baptist University.

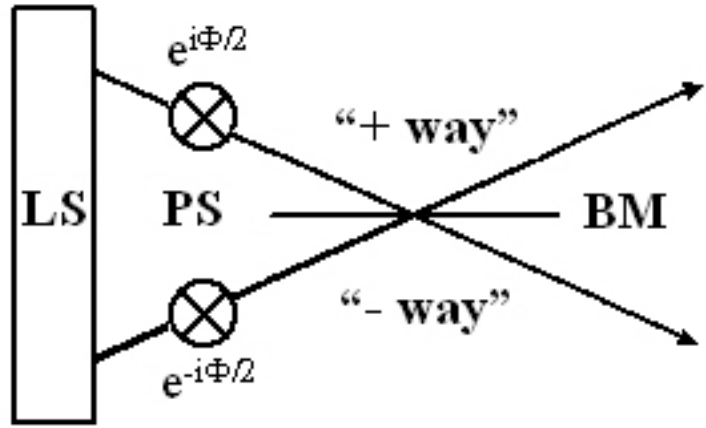

FIG. 1: Schematic two-way interferometer

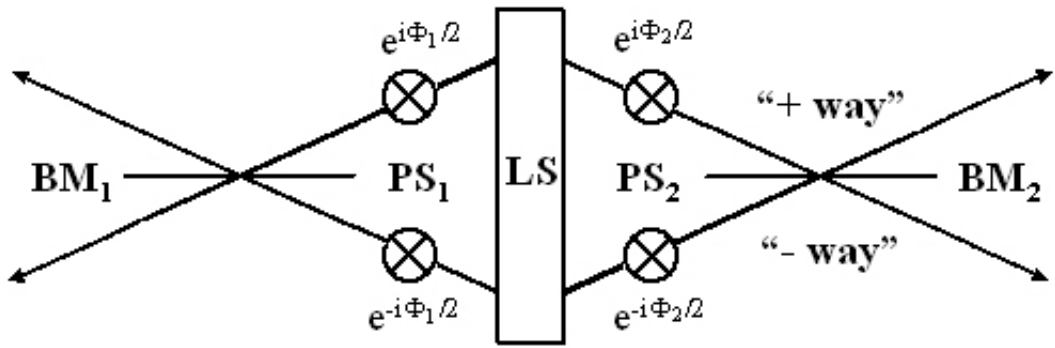

FIG. 2: Schematic two-particle four-beam interferometer

13